\documentclass[10pt,twocolumn,letterpaper]{article}
\usepackage{wacv}              % To produce the CAMERA-READY version
\usepackage{graphicx}
\usepackage{amsmath}
\usepackage{amssymb}
\usepackage{booktabs}
\usepackage{makecell}  % for \makecell command
\usepackage{multirow}
\usepackage{textcomp} % for \textsuperscript command
\usepackage{float}
\usepackage[table,xcdraw]{xcolor} % For row colors
\usepackage[pagebackref,breaklinks,colorlinks]{hyperref}

\usepackage[capitalize]{cleveref}
\crefname{section}{Sec.}{Secs.}
\Crefname{section}{Section}{Sections}
\Crefname{table}{Table}{Tables}
\crefname{table}{Tab.}{Tabs.}

\def\wacvPaperID{2100} % *** Enter the WACV Paper ID here
\def\confName{WACV}
\def\confYear{2025}

\begin{document}

%%%%%%%%% TITLE - PLEASE UPDATE
\title{STA-Unet: Rethink the semantic redundant for Medical Imaging Segmentation}

\author{Vamsi Krishna Vasa\\
Arizona State University\\
% Institution1 address\\
{\tt\small vvasa1@asu.edu}
% For a paper whose authors are all at the same institution,
% omit the following lines up until the closing ``}''.
% Additional authors and addresses can be added with ``\and'',
% just like the second author.
% To save space, use either the email address or home page, not both
\and
Wenhui Zhu \\
Arizona State University\\
% First line of institution2 address\\
{\tt\small wzhu59@asu.edu}
\and
Xiwen Chen \\
Clemson University\\
{\tt\small xiwenc@g.clemson.edu}
\and
Peijie Qiu \\
Washington University in St.Louis\\
{\tt\small peijie.qiu@wustl.edu}
\and
Xuanzhao Dong \\
Arizona State University\\
{\tt\small xdong64@asu.edu}
\and
Yalin Wang \\
Arizona State University \\
{\tt\small ylwang@asu.edu} 
}
\maketitle

%%%%%%%%% ABSTRACT
\begin{abstract}
In recent years, significant progress has been made in the medical image analysis domain using convolutional neural networks (CNNs). In particular, deep neural networks based on a U-shaped architecture (UNet) with skip connections have been adopted for several medical imaging tasks, including organ segmentation. Despite their great success, CNNs are not good at learning global or semantic features. Especially ones that require human-like reasoning to understand the context. Many UNet architectures attempted to adjust with the introduction of Transformer-based self-attention mechanisms, and notable gains in performance have been noted. However, the transformers are inherently flawed with redundancy to learn at shallow layers, which often leads to an increase in the computation of attention from the nearby pixels offering limited information. The recently introduced Super Token Attention (STA) mechanism adapts the concept of superpixels from pixel space to token space, using super tokens as compact visual representations. This approach tackles the redundancy by learning efficient global representations in vision transformers, especially for the shallow layers. In this work, we introduce the STA module in the UNet architecture (STA-UNet), to limit redundancy without losing rich information. Experimental results on four publicly available datasets demonstrate the superiority of STA-UNet over existing state-of-the-art architectures in terms of Dice score and IOU for organ segmentation tasks. The code is available at \url{https://github.com/Retinal-Research/STA-UNet}.
   
\end{abstract}

%%%%%%%%% BODY TEXT
\section{Introduction}
\label{sec:intro}

Leveraging advancements in deep learning, computer vision techniques have become integral to medical image analysis. Among these techniques, image segmentation holds significant importance. Specifically, precise and reliable segmentation of medical images is crucial, serving as a foundational component in computer-aided diagnosis and image-guided surgical procedures \cite{hatamizadeh2021unetrtransformers3dmedical, transunet}.

Current approaches to medical image segmentation predominantly utilize fully convolutional neural networks (FCNNs) with a U-shaped architecture \cite{unet, isensee2021nnu, Jin_2020}. The widely recognized U-Net \cite{unet}, a classic example of this architecture, features a symmetric Encoder-Decoder design linked by skip connections. The encoder extracts deep features with extensive receptive fields through multiple convolutional and down-sampling layers. The decoder then up-samples these deep features back to the original resolution for precise pixel-level semantic predictions, while the skip connections merge high-resolution features from various scales within the encoder to mitigate spatial information loss due to down-sampling. This well-crafted design has enabled U-Net to succeed significantly across numerous medical imaging tasks. The remarkable performance of these FCNN-based methods in cardiac segmentation, organ delineation, and lesion detection underscores CNNs' robust capability in learning distinguishing features.

While CNN-based techniques have demonstrated impressive results in medical image segmentation, they still fall short of the high accuracy standards required for clinical applications. Medical image segmentation remains a challenging problem, primarily due to the convolution operation's inherent focus on local features, making it difficult for CNNs to capture explicit global and long-range semantic interactions.
% Various approaches, such as atrous convolutional layers \cite{chen2017deeplabsemanticimagesegmentation,Gu_2019}, self-attention mechanisms \cite{schlemper2019attentiongatednetworkslearning, wang2018nonlocalneuralnetworks}, and image pyramids \cite{zhao2017pyramidsceneparsingnetwork}, have been proposed to address these shortcomings, but they still struggle with effectively modeling long-range dependencies. 
Recently, inspired by the success of Transformers in natural language processing (NLP) \cite{vaswani2023attentionneed}, researchers have started to explore their application in the vision domain \cite{carion2020endtoendobjectdetectiontransformers, dosovitskiy2021imageworth16x16words, liu2021swintransformerhierarchicalvision} to address the limitations of CNNs using self-attention. While the Vision Transformer (ViT) excels at capturing long-range dependencies across image patches with its large receptive field, it faces challenges in retaining fine-grained local context due to its lack of inherent locality.

To address this issue, recent approaches \cite{transunet, swin-unet, levit, hiformer} have proposed hybrid models that combine CNNs and ViTs in UNet architectures. However, these models significantly increase computational complexity and the number of parameters. Over-parameterization is a prevalent problem in deep learning, frequently resulting in feature redundancy and suboptimal feature representation \cite{dalvi2020analyzingredundancypretrainedtransformer, li2017pruningfiltersefficientconvnets, li2022self}. Despite its impact, the existing research has not thoroughly explored or considered this challenge. In addition to the methods discussed, several approaches aim to enhance the architectural design of UNet. For instance, Att-UNet \cite{atttentionUnet} introduces attention-based skip connections to filter out irrelevant features, while UNet++ \cite{zhou2018unetnestedunetarchitecture} replaces traditional skip connections, such as concatenation or addition, with nested dense skip pathways. UCTransNet \cite{wang2022uctransnetrethinkingskipconnections} provides an in-depth analysis of various skip connection strategies and proposes using a channel transformer as an alternative to conventional skip connections. Recently proposed Seg-Swinunet \cite{seg-swinunet} leverages the feature map with the highest semantic content (i.e., the decoder's final layer) to provide additional supervision to other blocks, reducing feature redundancy through feature distillation. However, we investigate this redundancy from a different perspective.

Our preliminary analysis indicates a significant similarity among blocks in the shallow layers of Transformer UNet architectures \cite{transunet, swin-unet, levit, hiformer}. This observation implies that the model exhibits a form of inertia learning pattern in the shallow layers, leading to a failure to effectively capture and encode complex contextual information. Existing research has seldom addressed this inherent limitation. \cite{vtsts} adapt the concept of superpixels from the pixel domain to the token domain, considering super tokens as a concise representation of visual information. This approach integrates sparse association learning, self-attention, and token space mapping to enhance visual token processing efficiency, leading to rich feature learning. In this study, we attempted to tackle redundancy by integrating Super Token Attention in UNet architecture and enhancing the performance of the multi-organ segmentation challenge. 

% In this study, we reported an empirical study conducted on Transformer UNet architecture \cite{transunet, swin-unet, levit, hiformer}, concluding that redundant features exist in the feature channel, with the shallow channels exhibiting more diversity than deep channels in a feature map. We attempted to mitigate this redundancy by utilizing the super tokens while enhancing the performance of UNet for multi-organ segmentation tasks.

% Super Token attention \cite{vtsts} has been tested for computer vision tasks such as classification, object detection and instance segmentation. The application of Super Token Attention (STA) in medical image segmentation remains an unexplored area in current research, which we aimed to bring to light for the first time. 

The main contributions of our work are three-fold: \textbf{(i)} We highlight the redundancy in the shallow layers of the transformer-based UNets to promote research in this area. \textbf{(ii)} We integrate the Super Token Attention (STA) block into the UNet architecture to minimize the redundancy observed in other Transformer-based UNet models while preserving the rich semantic information necessary for effective learning. \textbf{(iii)} Our comprehensive evaluation across four publicly available medical imaging datasets demonstrated the superiority of the proposed method over existing relevant SOTA methods in organ segmentation tasks.
% \textbf{(ii)} We present the concrete preliminary feature analysis to understand the redundancy seen in other methods and ablation study to understand the performance variations with respect to the STA block parameters. 

\section{Related Work}

\textbf{UNet-based architectures:} Early methods for medical image segmentation primarily relied on contour-based approaches and traditional machine learning techniques \cite{1194625, Held_1997}. However, the advent of deep convolutional neural networks (CNNs) brought significant advancements, with the introduction of Unet \cite{unet} specifically designed for medical image segmentation. The U-Net's distinctive U-shaped architecture, noted for its simplicity and exceptional performance, has inspired numerous variations, including Res-Unet \cite{8589312}, Dense-Unet \cite{li2018hdenseunethybriddenselyconnected}, U-Net++ \cite{zhou2018unetnestedunetarchitecture}, and UNet3+ \cite{huang2020unet3fullscaleconnected}. CNN-based architectures are flawed in capturing redundant information and do not focus on learning the dependencies between different regions of the canvas.  

\textbf{Transformer based UNet architectures :} Vaswani et al. \cite{vaswani2023attentionneed} introduced the Self-attention mechanism using Transformers in Natural Language Processing to weigh the importance of different words relative to each other. This advancement led to the development of the Vision Transformers (ViT) \cite{dosovitskiy2021imageworth16x16words,  liu2021swintransformerhierarchicalvision}, which adapts the transformer architecture to achieve comparable success in image processing tasks. These transformers are integrated with the UNet designs \cite{levit, swin-unet, transunet}, with the aim to combine the strengths of CNNs and Transformers. 

Chen et al. \cite{transunet} attempted to combine the Transformers in the encoder and decoder of the UNet architecture. The Transformer block in the encoder tokenizes image patches from a CNN feature map to capture global context. Meanwhile, the decoder upsamples these encoded features and merges them with high-resolution feature maps from the CNN. Although the ViT excels in capturing long-range dependencies between image patches (tokens) due to its large receptive field, it faces challenges in maintaining detailed local context because of its lack of inherent locality. To overcome this limitation, Swin-Unet \cite{swin-unet} adapts the attention mechanism using shifting window tokens \cite{liu2021swintransformerhierarchicalvision}. This allows the model to restrict window-based attention to local regions. Although this adaptation limits the redundancy, it is not completely eradicated from the shallow layers. Meanwhile, Zhu et al. \cite{seg-swinunet} proposed Seg-SwinUNet, which addresses performance issues in UNet for medical image segmentation by balancing supervision between the encoder and decoder and reducing feature redundancy. It tried to enhance UNet by using feature distillation to provide additional supervision from the most semantically rich feature map, improving segmentation accuracy with minimal computational overhead. However, the work is still limited to Swin-UNet, and no further study has been conducted to incorporate this approach with other architectures. 

Xu et al. \cite{levit} proposed LeViT-Unet, where LeViT \cite{graham2021levitvisiontransformerconvnets} as the encoder within the LeViT-UNet framework, as it effectively balances accuracy and efficiency in Transformer blocks. Additionally, skip connections integrate multi-scale feature maps derived from the Transformer and convolutional blocks of LeViT into the decoder. As the LeViT plays the central role in preserving and passing information to the decoder, the redundant token information couldn't be avoided, Leading to increased computational cost. Recently proposed Hiformer \cite{hiformer} integrates CNN and transformer architectures to capture both local and global features for medical image segmentation. It employs multi-scale feature representations using a Swin Transformer and CNN-based encoder, combined through a Double-Level Fusion (DLF) module in the encoder-decoder structure. Extensive experiments show HiFormer’s superior performance in accuracy and efficiency compared to other methods.

\textbf{Mitigating feature redundancy:} Oktay et al. \cite{atttentionUnet} proposed Attention Gates (AG) to focus on target structures of varying shapes and sizes by suppressing irrelevant regions and highlighting important features. This eliminates the need for external localization modules, as AGs can be easily integrated into CNN architectures like U-Net with minimal computational cost. Zhou et al. \cite{zhou2018unetnestedunetarchitecture} proposed UNet++ architecture to deeply supervise the encoder-decoder network that connects the encoder and decoder through nested, dense skip pathways. These redesigned pathways aim to reduce the semantic gap between the encoder's and decoder's feature maps, making the learning task easier for the optimizer. Wang et al. \cite{wang2022uctransnetrethinkingskipconnections} proposed UCTransNet replaces traditional U-Net skip connections with the Channel Transformer (CTrans) module, which includes two sub-modules: the Channel Cross fusion with Transformer (CCT) for multi-scale channel fusion and Channel-wise Cross-Attention (CCA) to guide fused features into the decoder. This new connection structure addresses semantic gaps between encoder and decoder features for improved segmentation. Zhu et al. \cite{seg-swinunet} proposed model balances supervision between the encoder and decoder and reduces feature redundancy in UNet by providing additional supervision from the most semantically rich feature map (the last layer of the decoder) to other blocks. It leverages feature distillation to minimize redundant information and enhance learning efficiency. This method integrates seamlessly into existing UNet architectures with minimal computational overhead, improving performance across various medical image segmentation tasks.

%Although these models significantly improved the performance of the medical imaging segmentation, they lack in obtaining efficient and effective global representations at the early stages of a neural network. To address this problem Huang et al. \cite{vtsts} proposed Super Token Attention (STA), inspired by Superpixels \cite{jampani2018superpixelsamplingnetworks}. STA mechanism first predicts super tokens through sparse associations and then applies self-attention in the super token space. This approach significantly reduced computational complexity while effectively capturing global context compared to traditional token-based self-attention. We attempted to effectively integrate the concept with Unet architecture to perform Medical Imaging Segmentation task

\section{Preliminary Analysis}
\label{premiliminary analysis}
Here, we employ Centered Kernel Alignment (CKA) \cite{kornblith2019similarity} to investigate the recent popular U-net models, including SwinUnet, LeViT-Unet, TransUnet, and HiFormer. This technique allows us to compute the block-wise similarity even when the layers' sizes are different. 
The block-wise similarity matrices are able to offer insights into how different neural network architectures learn and represent information at various layers (or blocks) throughout the training process. 

Mathematically, given two sets of representations $\boldsymbol{X}$ and $\boldsymbol{Y}$, we first compute their gram matrices $\boldsymbol{K},\boldsymbol{L}$ via the Radial Basis Function (RBF),
\begin{align}
    % \boldsymbol{K}_{ij} &= \exp\left(- \gamma_{\boldsymbol{x}}  \|\boldsymbol{X}_i - \boldsymbol{X}_j\|^2 \right), 
    % \\ \nonumber
    % \boldsymbol{L}_{ij} &= \exp\left(-\gamma_{\boldsymbol{y}} \|\boldsymbol{Y}_i - \boldsymbol{Y}_j\|^2\right),
        \boldsymbol{K}_{ij} &= \exp\left(-  \|\boldsymbol{X}_i - \boldsymbol{X}_j\|^2 \right), 
    \\ \nonumber
    \boldsymbol{L}_{ij} &= \exp\left(- \|\boldsymbol{Y}_i - \boldsymbol{Y}_j\|^2\right),
\end{align}
where $\boldsymbol{K}_{ij}$ and $\boldsymbol{L}_{ij}$ denotes the $i$th row and $j$th column element. $\boldsymbol{X}_i, \boldsymbol{X}_j, \boldsymbol{Y}_i - \boldsymbol{Y}_j$ denotes the $i$th and $j$th sample of the set $\boldsymbol{X},\boldsymbol{Y}$, respectively. 
% We set $\gamma_{\boldsymbol{x}} = \frac{1}{d_{\boldsymbol{X}}}$ and $\gamma_{\boldsymbol{y}} = \frac{1}{d_{\boldsymbol{Y}}}$ as default. 
% Here, $d_{\boldsymbol{X}},d_{\boldsymbol{Y}}$ denote the dimension of the sample for $\boldsymbol{X},\boldsymbol{Y}$, respectively. 
Afterward, the similarity matrix is computed through RBF-CKA as,
\begin{align}
   RBF-CKA= \frac{\text{tr}(\boldsymbol{KHLH})}{\sqrt{\text{tr}(\boldsymbol{KHKH}) \, \text{tr}(\boldsymbol{LHLH})}},
\end{align}
where $\boldsymbol{H}_n = \boldsymbol{I}_n - \frac{1}{n} \mathbf{1} \mathbf{1}^\top$ denotes the centering matrix. $ \boldsymbol{I}_n$ denotes a identity matrix with shape $n\times n$, where $n$ denotes the number of samples in the set.

\begin{figure}[H]
  \includegraphics[width=\columnwidth]{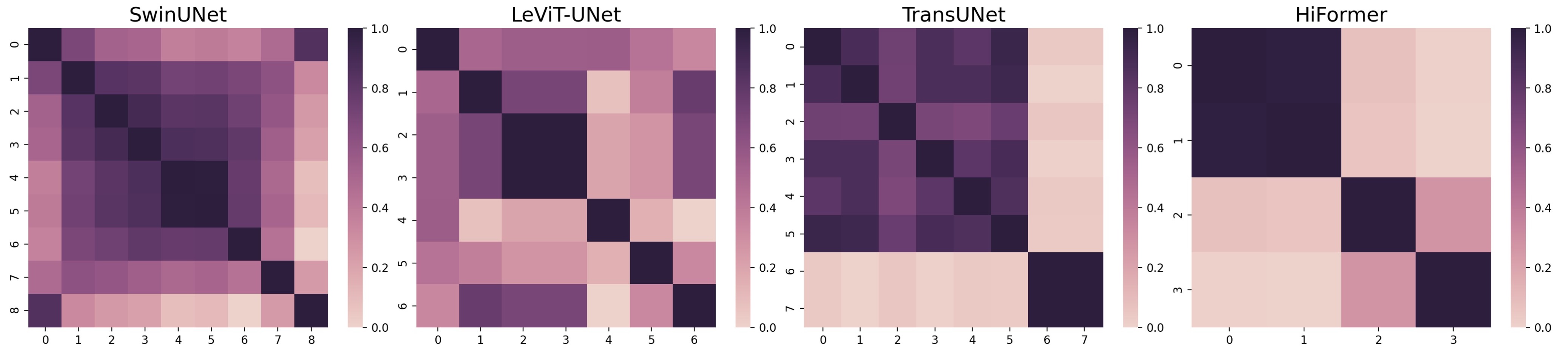}  % Replace with your figure file and remove the example
  \caption{The block-wise similarity calculated by RBF-CKF \cite{kornblith2019similarity}. The indices are ordered from shallow blocks to deep blocks. For the sake of better visualization, we normalize it to 0$\sim$1 by using min-max normalization. }
  \label{fig:CKF}
\end{figure}

In Fig.~\ref{fig:CKF}, we present the results of our investigation, where high values in off-diagonal elements typically indicate strong similarities between corresponding blocks. Across various architectures, the similarity matrices generally exhibit higher degrees of similarity among blocks in the shallow blocks, suggesting a high level of redundancy in these early layers. This immediately implies that the neural network is lazy at the shallow blocks and fails to learn rich information.
This observation motivates our efforts to address and reduce such redundancy in the proposed work.

% \begin{figure*}[!t]
%     \centering
%     \includegraphics[width=\columnwidth]{images/draw.jpg}
%     \caption{The block-wise similarity calculated by CKF \cite{kornblith2019similarity}. The indices ordered from shallow blocks to the deep blocks.}
%     \label{fig:CKF}
% \end{figure*}

\section{Method}
\subsection{Super Token Attention Module} \label{sta module section}
% You need to explain what a super token is and why it is needed before introducing unet, giving some schematics and mathematical expressions (This previous analysis should be relevant, you need to connect them, a clear intuition is very important). This is very important. Only after that you need to introduce Unet.
Based on the analysis presented in Section \ref{premiliminary analysis}, there is clear evidence of redundancy in the shallow layers of transformer-based architectures, which results in inefficient information retention. Super tokens \cite{vtsts} can mitigate this flaw by learning efficient global representation. Super tokens are considered as a concise representation of visual information by adapting the concepts of superpixels \cite{jampani2018superpixelsamplingnetworks} from pixel domain to token domain. This method combines sparse association learning, self-attention, and token space mapping to improve the efficiency of visual token processing. We re-introduce the Super Token Attention in form of STA Module in the UNet architecture to leverage its benefits for the Medical Image Segmentation task.

% Huang et al. \cite{vtsts} introduces the Super Token attention to learn efficient global representation inspired by the idea of Superpixels \cite{jampani2018superpixelsamplingnetworks}. Superpixels perform image over-segmentation by clustering visually similar pixels, simplifying the image into fewer segments, which facilitates more efficient processing in later stages. \cite{vtsts} adapt the concept of superpixels from the pixel domain to the token domain, considering super tokens as a concise representation of visual information. This approach integrates sparse association learning, self-attention, and token space mapping to enhance visual token processing efficiency. We re-introduce the Super Token Attention in form of STA-block in the UNet architecture to leverage its benefits for the Medical Image Segmentation task. 

Huang et al. achieves Super Token Attention in three stages, Convolutional Position Embedding (CPE). Super Token Attention (STA) and Convolutional Feed-Forward Network (ConvFFN). The CPE stage comprises of the Residual DW layer with $3\times3$ depth-wise convolution over input $X_{in} \in \mathbb{R}^{H \times W \times C}$ . CPE is more flexible for arbitrary input resolutions as it can learn absolute positions through zero padding, unlike absolute \cite{vaswani2023attentionneed} and relative \cite{liu2021swintransformerhierarchicalvision, li2022mvitv2improvedmultiscalevision} positional encodings. Emphasizing on ConvFFN, which is a collection of convolutional layers with GeLu \cite{hendrycks2023gaussianerrorlinearunits} activation to calculate the learned representation post attention mechanism. The presence of skip connection compensate for this these blocks with no performance degradation, Thus we restrict this stage in our STA-block:

\setlength{\abovedisplayskip}{0cm}
\setlength{\belowdisplayskip}{0cm}

\begin{align}
\boldsymbol{X} = \text{CPE}(X_{\text{in}}) + X_{\text{in}} 
\end{align}

\begin{align}
    \boldsymbol{Y} = \text{STA}(\text{LN}(\boldsymbol{X})) + \boldsymbol{X}
\end{align}

\begin{figure}[!t]
  \centering
  \includegraphics[width=\columnwidth]{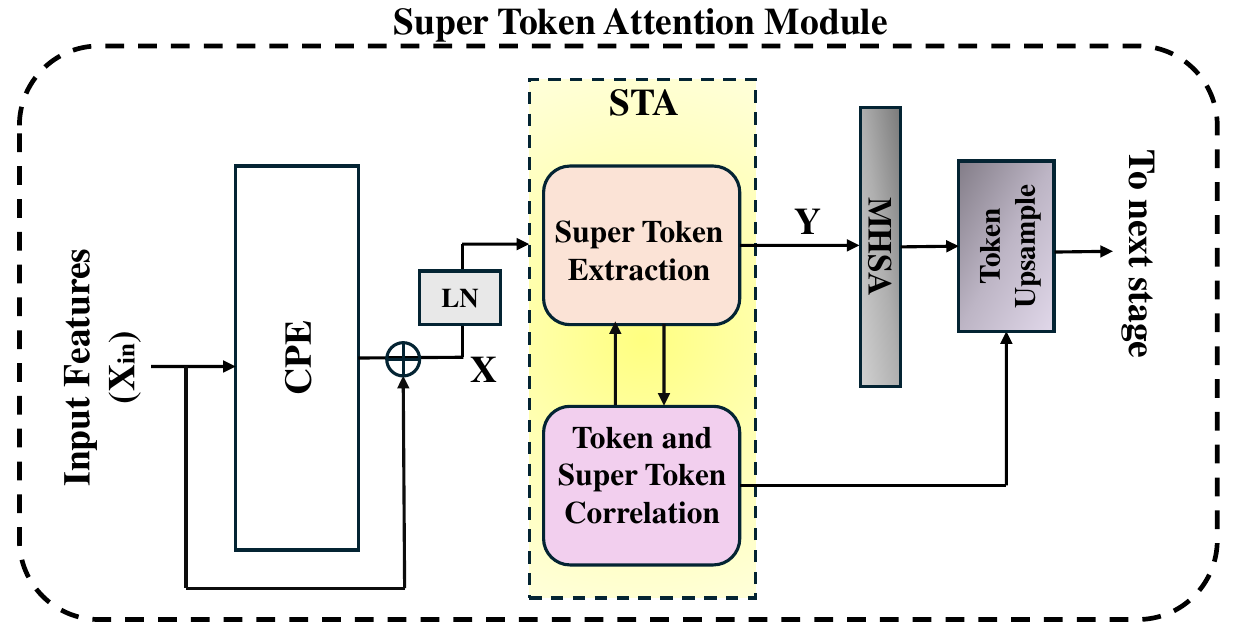}  % Replace with your figure file and remove the example
  \caption{Super Token Attention (STA) Module incorporated in the UNet architecture.}
  \label{sta module}
  \vspace{-0.5cm}
\end{figure}

The revised STA Module is illustrated in Fig \ref{sta module}. In STA stage, Super Token Extraction follows k-mean based superpixel algorithm in Super Sampling Networks \cite{jampani2018superpixelsamplingnetworks} to map the pixels to token space.
% The tokens ($X \in \mathbb{R}^{C \times N}$, where $N$ is the number of tokens given by $H \times W$)
The tokens can be denoted as $X \in \mathbb{R}^{C \times N}$. N is the number of tokens given by $H \times W$. Each token (denoted by $X_{i} \in \mathbb{R}^{C \times 1}$) belongs to one of the Super tokens (denoted by $S \in \mathbb{R}^{mxc})$. $m$ is the number of super tokens. If the grid size of $h \times w$, number of super tokens $m$ can be calculated as $m = \frac{H}{h} \times \frac{W}{w}$. 

\begin{figure*}[!t]
  \centering
  \includegraphics[width=\textwidth]{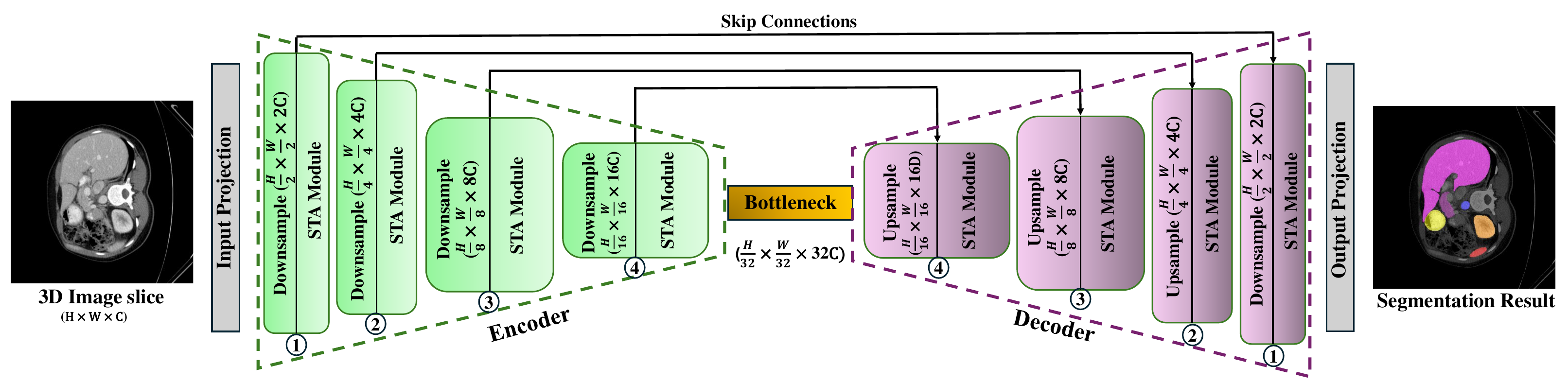}  % Replace with your figure file and remove the example
  \caption{Pictorial representation of the proposed STA-Unet architecture. The number in the circle denote the stage number.}
  \label{architecture}
  \vspace{-0.5cm}
\end{figure*}

Token and Super token correlation aims to calculate the $X_{i}$ to $S_{j}$ association ($Q_{ij}$) and update the Super tokens. We calculate the $Q$ using Eq. \ref{Eq 5} to attain the the attention type weights.  

\begin{align}
\label{Eq 5}
Q = Softmax(\frac{XS}{\sqrt{C}})
\end{align}

We limited this to a single iteration based on conclusive experiments and performance analysis. The super token is then updated as weighted sum of tokens defined as:

\begin{align}
S = (\bar{Q})^{T}X
\end{align}

For reduced computational power, we limit the association calculations to the surrounding nine super tokens. We apply the standard self-attention to the sampled super tokens $S \in \mathbb{R}^{m \times C}$:

\begin{align}
    Attn(S) = Softmax(\frac{q(S)k^{T}(S)}{\sqrt{C}})v(S)
\end{align}

$q(S) = SW_{q}$, $k(S) = SW_{k}$, $v(S) = SW_{v}$, where ($W_{q}, W_{k}, W_{v}$) are the linear function parameters.   
Lastly, We Upsample the Tokens to introduce the lost local features using the association map $Q$:

\begin{align}
    Upsample(Attn(S) = QAttn(S)
\end{align}

\begin{figure}[!t]
  \centering
  \includegraphics[width=\columnwidth]{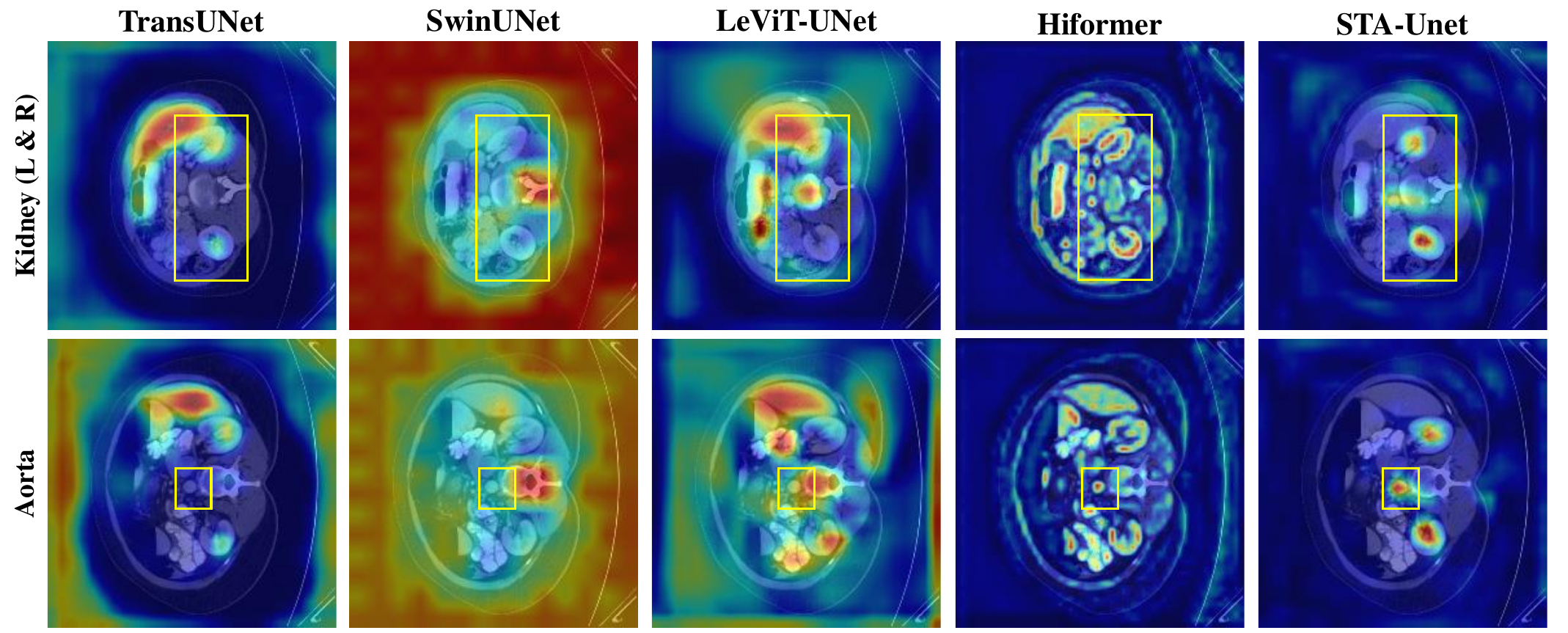}  % Replace with your figure file and remove the example
  \caption{Attention maps from Decoder (Stage-4) Layers for various transformer based UNet architectures.}
  \label{attention maps panel}
  \vspace{-0.5cm}
\end{figure}

\begin{table}[H]
\centering
\resizebox{\columnwidth}{!}{%
\begin{tabular}{cccc}
\hline
\textbf{Encoder} \textbackslash \textbf{Decoder} & \textbf{No. of layers} & \textbf{Token size} & \textbf{No. of Heads} \\ \hline
Stage - 1        & 1                      & 16 x 16             & 2                     \\
Stage - 2        & 2                      & 8 x 8               & 4                     \\
Stage - 3        & 3                      & 4 x 4               & 8                     \\
Stage - 4        & 4                      & 2 x 2               & 16                    \\ \hline
\end{tabular}%
}
\caption{The optimal parameters for Super Token Attention (STA) blocks in the each stage of Encoder/Decoder.}
\label{sta_dims}
\end{table}

The Multi-head setting is not included for clear understanding. The pseudocode for Super Token Attention can be found in the supplementary material of \cite{vtsts}. The parameter values for the STA blocks for each stage of Encoder and Decoder in UNet are reported in Table \ref{sta_dims}. We discuss the relationship between performance and choice of parameter values in Section \ref{ablation study}.

We also visualize the attention maps obtained from the existing transformer-based UNet architectures in Fig. \ref{attention maps panel}. It is worth mentioning that Super Token Attention has precisely given higher weights to the different regions of interest when classifying smaller organs such as Kidneys and Aortas in shallow blocks. 

\subsection{STA-UNet architecture}

We dedicate this section to providing a brief overview of the proposed Unet architecture (same is illustrated in Fig \ref{architecture}). Similar to any other UNt architecture, the proposed model comprises Encoder, Decoder, Bottleneck, and Skip connections. The key performance enhancer in this architecture is the Super Token Attention (STA) modules integrated at each stage of the encoder and decoder. On the contrary, with the Transformer Blocks leveraged in the latest UNet architectures \cite{transunet, swin-unet, levit}, we implement dimensional changes in the convolution layers, filtering essential information before applying the attention mechanism. The input image is downsampled to the half dimensions ($H/2 \times W/2$) and channel (C)-dimension is doubled at each stage. The positional embeddings are extracted in the STA Module (in the CPE stage), followed by Super Tokens generation and calculating the correlation between Tokens and Super Tokens as discussed in Section \ref{sta module section}. Symmetrical Decoder architecture is adapted with the combination of Upsample block (to obtain the original image shape) and STA modules. The context features extracted during the processing are concatenated with multi-stage features from the encoder through skip connections. This fusion mitigates the loss of spatial information typically incurred by down-sampling, thereby enhancing the model's ability to retain fine-grained details. Detailed information for each component is documented in this section.

\noindent\textbf{Encoder} 
\label{encoder}
The Encoder consists of two components in each stage. The Downsample block is followed by an STA module. The Downsample block deals with the dimensionality reduction in the forward propagation as we are not relying on the patch merging stage \cite{swin-unet}. The Downsampling block is made up of 2 layers of Convolution and Batch Normalization. The convolution layer possesses the kernel size of $3\times3$ and the stride of 2 with padding set to 1. We then process the output features with ReLU \cite{relu} activation and reduce the dimensions using the Maxpooling layer. To retain the complete spatial information from this stage, we pass the output from ReLu activation to the decoder through skip connections, unlike the traditional Unet architecture.

\noindent\textbf{Decoder}
Drawing inspiration from \cite{unet}, the Decoder is designed to be symmetrical to the encoder. Decoder also consists of two components, i.e., the Upsample block and the STA module. Upsample block consists of a ConvTranspose2d layer to increase the spatial dimensions of the input features, followed by the convolution layer similar to the one discussed in Encoder (Section \ref{encoder}). We concatenate the feature map from the encoder passed through the skip connection with the features obtained from the previous decoder stage (output from the STA module). The complete spatial information (before Maxpooling in Encoder) helps us present the spatial information along with the contextual features from the attention mechanism for improved learning. We then input the resultant to the Upsample block. The Decoder then passes the output to the Output projection layer to obtain the input image dimensionality and the channel size equal to the segmentation classes. The output is processed through the Softmax layer along the C-axis to obtain the class probabilities. 

\noindent\textbf{Bottleneck} comprises 2 Convolution layers followed by the BatchNorm layer. The ReLU activation function is applied to the output. 
% We tabulated the parameter values used to design the Encoder/Decoder STA blocks in Table. ~\ref{sta_dims}. We will discuss these parameters at length in \ref{sta_block}. These parameters are finalized after analyzing the performance over multiple experimental iterations. To get the multi-channeled output for each class in the segmentation task, We used the output layer comprising of 2d Convolution layer with the output channels equivalent to the number of classes. The same is followed by the Softmax layer along C-dimension to obtain the class probabilities. 

% Similar to \cite{unet}, \textbf{Skip connections} are used to preserve the spatial information and facilitate the gradient flow. In the recent derivations of Unet, the output features obtained from each encoder level are passed to the corresponding counter level of the decoder. Instead of the contextual features derived from the STA blocks, we concatenated the feature map obtained from the convolution layer (before Max pooling) in the Stage-1 of the Encoder levels with the previous decoder level's output. The same is showcased in the Right part of Fig. \ref{architecture}.

\section{Experiments and results}
\subsection{Datasets}

We validated the effectiveness of the proposed method on four publicly available datasets: Synapse Multi-Organ Segmentation, Automated cardiac diagnosis challenge (ACDC) dataset \cite{8360453}, Nuclear segmentation (MoNuSeg) \cite{8880654, kumar2017dataset}, and Gland segmentation in Colon Histology Images (GlaS) \cite{sirinukunwattana2016glandsegmentationcolonhistology}. Following \cite{swin-unet}, \cite{transunet}, \cite{hiformer}, we trained the proposed method using the Synapse Multi-Organ Segmentation dataset. The dataset includes 30 cases, encompassing a total of 3,779 axial abdominal CT images. The segmentation masks are provided for 13 abdominal organs, out of which we used 9 classes for training the proposed model. For model development, 18 cases are allocated for training, while 12 cases are designated for testing. Performance is assessed based on the segmentation of eight abdominal organs, with the Average Dice Similarity Coefficient (DSC) used as the primary evaluation metric.  The ACDC dataset consists of 100 cardiac MRI scans from a diverse patient cohort, with annotations for the left ventricle (LV), right ventricle (RV), and myocardium (Myo). In line with prior work \cite{transunet, 10030763}, we partition the dataset into 70 cases (1,930 axial slices) for training, 10 cases for validation, and 20 cases for testing. The performance of our method is evaluated using the Dice Similarity Coefficient (DSC) as the metric. The GlaS \cite{gland-seg} and MoNuSeg \cite{Monuseg} datasets are the collection of microscopic images. The GlaS dataset contains 85 images designated for training and 80 images for testing. The MoNuSeg dataset includes 30 images for training and 14 images for testing. The performance of the latter two datasets is evaluated using the average Dice Similarity Coefficient (DSC) and Intersection over Union (IoU) as metrics.

\subsection{Implementation details}

We followed the straightforward training regime for easy reproducibility mehtods~\cite{swin-unet,transunet,uctransunet,seg-swinunet}. The Synapse CT dataset consists of 3D CT scans with each slice mapped in the Grayscale domain. To train on this dataset, we extracted each slice and center-cropped it to retain the 224 x 224 image for the input. We trained our model for 300 epochs using a Stochastic Gradient Descent (SGD) optimizer for smoother convergence. The batch size was set to 8. The initial learning rate \(lr_{initial}\) was set to be \(1 \times 10^{-2}\). The learning rate for each iteration \(lr_{t}\) of the epoch is determined by the Eq. \ref{lr}. Where t denotes the current iteration, N denotes the maximum number of iterations in one epoch.
% \vspace{-1em}  % Adjust this value as needed
\begin{equation}
\text{lr}_{t} = \text{lr}_{\text{initial}} \times \left(1.0 - \frac{t}{N}\right)^{0.9}
\label{lr}
% \vspace{-0.5em}  % Adjust this value as needed
\end{equation}
We trained the model to converge on the summation of Cross-Entropy and Dice loss, maintaining the weights of 0.4 and 0.6, respectively. To tackle the limited dataset problem, we incorporated the following data augmentation traits: Random flips (Horizontal) and rotations with a probability of 0.5. We followed the same experimental setup for the ACDC dataset. For GlaS and MoNuSeg datasets, the batch size was 18. We used the initial learning rate of \(1 \times 10^{-3}\) and updated the learning rate using Cosine Scheduler. We utilized the computational power of Nvidia RTX3090 with 24G memory to conduct our experiments. We compare our methods with recent SOTA models, including UNet \cite{unet}, R50 U-Net~\cite{Diakogiannis_2020}, Att-UNet~\cite{atttentionUnet}, TransUNet~\cite{transunet}, SwinUNet~\cite{swin-unet}, LeViT-UNet~\cite{levit}, HiFormer~\cite{hiformer}, and Seg-SwinUNet\cite{seg-swinunet}.

% We evaluate the effectiveness of our proposed loss on both SwinUNet and UNet in these four datasets, and the training setups (i.e., batch size, optimizer, learning rate, etc.) are consistent with \cite{swimUNet}. All experiments were conducted with an input image size of 224 x 224 and the same data augmentation and preprocessing in~\cite{hiformer,pvt-cascade,uctransUNet}, using an Nvidia GTX3090 with 24GB of memory for training. Following in~\cite{swimUNet,UNet}, pre-trained weights from ImageNet were employed in SwinUNet, while UNet was trained from scratch.

% \noindent\textbf{Comparison with state-of-the-art methods.} We compare our methods with recent SOTA models, including R50 UNet~\cite{transUNet}, Att-UNet~\cite{atttentionUNet}, UNet++~\cite{UNet++}, TransUNet~\cite{transUNet},swinUNet~\cite{swimUNet}, LeViT-UNet~\cite{letViT}, DeepLabv3~\cite{deeplabv3}, HiFormer~\cite{hiformer}, PVT-cascade~\cite{pvt-cascade}, UCTransNet~\cite{uctransUNet}, MedT~\cite{medT}. 

\subsection{Ablation Study}
\label{ablation study}
Understanding the impact of individual parameters on model performance is crucial for determining the optimal architecture. To gain insights into the effects of varying Token size and Attention heads in our proposed model, we conducted an ablation study on the GlaS dataset. 

\begin{table}[H]
\centering
\caption{Performance of STA-UNet with different Attention heads at each stage. The $\rightarrow$ indicates a level change in Encoder, and the reverse trend is followed with the token sizes in Decoder.}
\label{tab:att-head}
\vspace{-0.5em} % Adjust vertical space before the table
\begin{tabular}{ccc}
\hline
\textbf{\begin{tabular}[c]{@{}c@{}}Attention heads \\ (from Stage 1 to 4)\end{tabular}} & \textbf{\begin{tabular}[c]{@{}c@{}}Mean Dice\\ (\%)\end{tabular}} & \textbf{\begin{tabular}[c]{@{}c@{}}IOU\\ (\%)\end{tabular}} \\ \hline
8 $\rightarrow$ 16 $\rightarrow$ 32 $\rightarrow$ 64                                    & 89.91                                                             & 76.95                                                       \\
4 $\rightarrow$ 8 $\rightarrow$ 16 $\rightarrow$ 32                                     & 90.41                                                             & 83.27                                                       \\
\rowcolor[HTML]{FCFF2F} 
2 $\rightarrow$ 4 $\rightarrow$ 8 $\rightarrow$ 16                                      & 91.03                                                             & 84.29                                                       \\
1 $\rightarrow$ 2 $\rightarrow$ 4 $\rightarrow$ 8                                       & 90.56                                                             & 83.52                                                       \\ \hline
\end{tabular}
\end{table}

\begin{figure}[H]

  \includegraphics[width=\columnwidth]{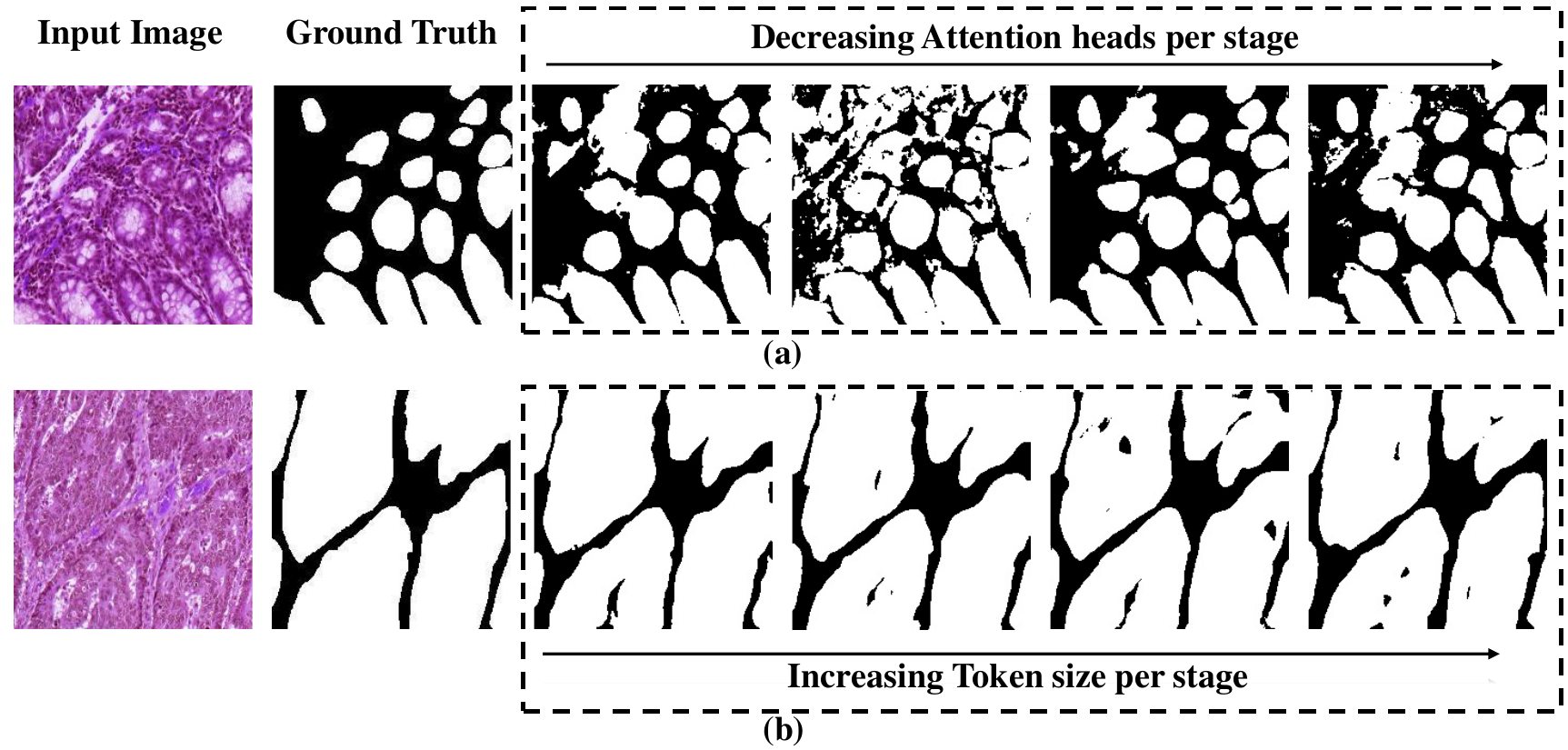}  % Replace with your figure file and remove the example
  \caption{Illustration of Ablation studies on Glas dataset. (a) Decreasing the attention heads leads to the accurate segmentation of Glands. (b) Increasing the Token size leads to indistinguishable changes.}
  \label{ab res}
\end{figure}

In transformer-based architectures, increasing attention heads often leads the model to capture information from more regions and determine the importance with respect to the decision. It is evident from Table \ref{tab:att-head} that the super token attention mechanism can achieve superior performance with limited attention heads, bringing down the overlap in focus or redundancy. The same is illustrated in Fig. \ref{ab res} (a). We chose the highlighted Attention heads trend based on the performance of our proposed method. 

The Token size impacts the model's ability to capture spatial details and contextual information. Larger tokens provide broader context but can reduce resolution, while smaller tokens capture finer details but may increase computational complexity. The same is evident from Table \ref{tab:tok-size}. Balancing token size is crucial for optimizing both model performance and efficiency. 

\begin{table}[H]
\centering
\caption{Performance of STA-UNet with different Token sizes at each stage. The $\rightarrow$ indicates a level change in Encoder, and the reverse trend is followed with the token sizes in Decoder.}
\label{tab:tok-size}
\resizebox{\columnwidth}{!}{%
\begin{tabular}{cccc}
\hline
\textbf{\begin{tabular}[c]{@{}c@{}}Token size \\ (from Stage 1 to 4)\end{tabular}} &  \textbf{FLOPs}                                       & \textbf{Mean Dice} & \textbf{IOU}   \\ \hline
32 $\rightarrow$ 16 $\rightarrow$ 8 $\rightarrow$ 4 & $59.02\times10^{3}M$        & 90.8               & 83.80          \\
\rowcolor[HTML]{FFFE65} 
16 $\rightarrow$ 8 $\rightarrow$ 4 $\rightarrow$ 2 & $60.30\times10^{3}M$ & 91.03     & 84.29 \\
8 $\rightarrow$ 4 $\rightarrow$ 2 $\rightarrow$ 1 & $67.92\times10^{3}M$          & 90.80              & 83.81          \\
4 $\rightarrow$ 2 $\rightarrow$ 1 $\rightarrow$ 1 & $91.95\times10^{3}M$           & 90.31              & 83.76          \\ \hline
\end{tabular}%
}
\end{table}

\begin{figure*}[!ht]
  \centering
  \includegraphics[width=\textwidth]{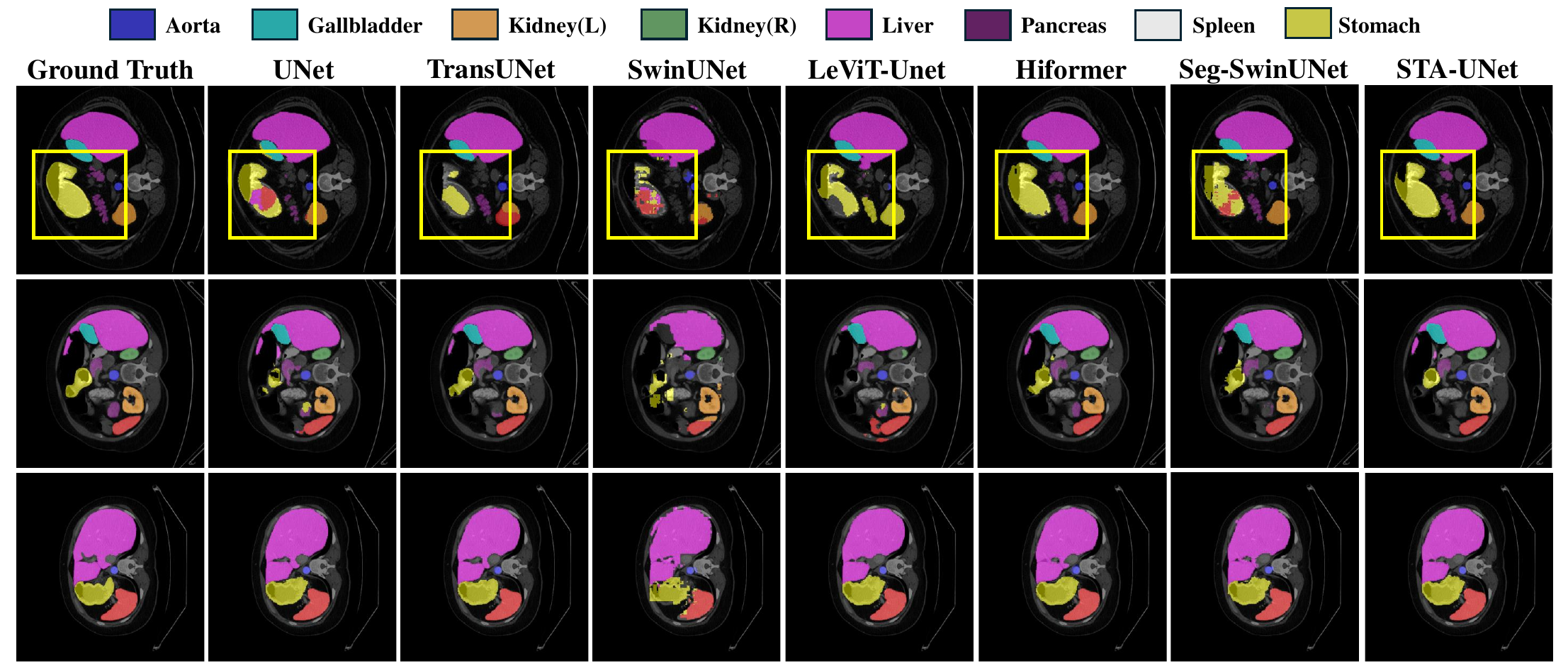}  % Replace with your figure file and remove the example
  \caption{Comparison of segmentation performance in Synapse dataset with Transformer-based UNet architectures. \textcolor{yellow}{Yellow} box highlights how the baseline methods handled Pancreas and Spleen segmentation.}
  \label{synapse results}
\end{figure*}

\begin{table*}[ht]
\centering
\caption{Comparison with SOTA methods on Synapse multi-organ CT dataset. $\Delta_{UNet}$ denotes the improvement gain (\%) by comparing with the U-Net \cite{unet}. $\Delta_{TransUNet}$ denotes the improvement gain (\%) by comparing with the TransUNet \cite{transunet}}
\label{tab:synapse}
\resizebox{\textwidth}{!}{%
\begin{tabular}{l|c|cccccccc}
\toprule
\textbf{Methods} & \textbf{Average DSC} &  \textbf{Aorta} & \textbf{Gallbladder} & \textbf{Kidney(L)} & \textbf{Kidney(R)} & \textbf{Liver} & \textbf{Pancreas} & \textbf{Spleen} & \textbf{Stomach} \\ \hline
R50 U-Net & 74.68 & 87.74 & 63.66 & 80.60 & 78.19 & 93.74 & 56.90 & 85.87 & 74.16 \\
R50 Att-UNet  & 75.57 & 55.92 & 63.91 & 79.20 & 72.71 & 93.56 & 49.37 & 87.19 & 74.95 \\
Att-UNet & 77.77 & 89.55 & 68.88 & 77.98 & 71.11 & 93.57 & 58.04 & 87.30 & 75.75 \\
U-Net & 76.85 & 89.07 & 69.72 & 77.77 & 68.60 & 93.43 & 53.98 & 86.67 & 75.58 \\
\hline
TransUNet  & 77.48 & 87.23 & 63.13 & 81.87 & 77.02 & 94.08 & 55.86 & 85.08 & 75.62 \\
SwinUNet  & 79.13  & 85.47 & 66.53 & 83.28 & 79.61 & 94.29 & 56.58 & 90.66 & 76.60 \\
LeViT-UNet & 78.53  & 78.53 & 62.23 & 84.61 & 80.25 & 93.11 & 59.07 & 88.86 & 72.76 \\
HiFormer  & 80.29 & 85.63 & 73.29 & 82.39 & 64.84 & 94.22 & 60.84 & 91.03 & 78.07 \\
Seg-SwinUNet & 80.54 & 86.07 & 69.65 & 85.12 & 82.58 & 94.18 & 61.08 & 87.42 & 78.22 \\
% HiFormer - B & 80.39 & 14.7 & 86.21 & 65.69 & 85.23 & 79.77 & 94.61 & 59.52 & 90.99 & 81.08 \\
% HiFormer - L & 80.69 & 19.14 & 87.03 & 68.61 & 84.23 & 78.37 & 94.07 & 60.77 & 90.44 & 82.03 \\ 
\hline
\rowcolor{blue!20}
STA-Unet & \textbf{80.69} & 89.10 & 68.34 & 84.97 & 79.44 & 93.39 & 63.32 & 88.69 & 78.26  \\ 
\rowcolor{blue!20}
$\Delta_{UNet}$ & \textcolor{blue}{+4.99} & \textcolor{blue}{+3.00} & \textcolor{red}{-1.97} & \textcolor{blue}{+9.25} & \textcolor{blue}{+15.80} & \textcolor{red}{-0.43} & \textcolor{blue}{+17.30} & \textcolor{blue}{+2.33} & \textcolor{blue}{+3.54}\\
\rowcolor{blue!20}
$\Delta_{TransUNet}$ & \textcolor{blue}{+4.14} & \textcolor{blue}{+2.14} & \textcolor{blue}{+8.25} & \textcolor{blue}{+3.70} & \textcolor{blue}{+3.10} & \textcolor{red}{-0.73} & \textcolor{blue}{+13.35} & \textcolor{blue}{+4.24} & \textcolor{blue}{+3.49}\\
\hline
\end{tabular}%
}
% \vspace{-0.8cm}
\end{table*}

But in the proposed Super Token Attention approach, we notice that the performance has very minute changes with the change of token size, Thus limiting the dependency of performance over the Token size; the same is evident from Fig. \ref{ab res} (b). Based on the study, we chose the model with relatively lower Floating Point Operations per second to lower the computational complexity.

\begin{figure*}[!ht]
  \centering
  \includegraphics[width=\textwidth]{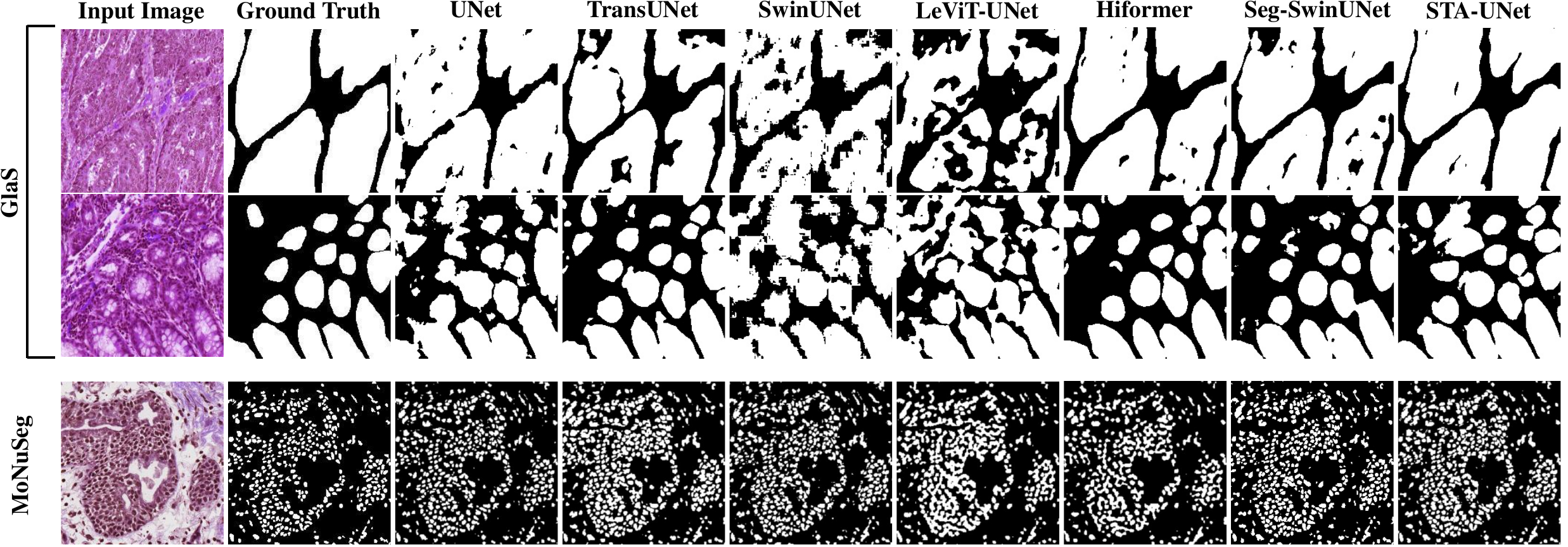}  % Replace with your figure file and remove the example
  \caption{Comparison of segmentation performance on GlaS (Glands) and MoNuSeg (Nuclear) Dataset.}
  \label{Glas results}
\end{figure*}

\subsection{Results}

The performance analysis is reported in Table~\ref{tab:synapse} for the Synapse dataset, Table~\ref{tab:acdc} for the ACDC dataset, and Table~\ref{tab:glas} for Glas and MoNuSeg datasets. Our main conclusion is that our proposed architecture is effective and computationally rational and achieved significant improvement over quantitative metrics. We reported the gain/loss in percent with respect to the UNet \cite{unet} and the first-ever proposed transformer-based UNet architecture, TransUNet \cite{transunet}. We observe substantial improvements of 4.99\%, 2.86\%, 6.53\%, and 6.03\% across the four datasets when compared to UNet. Similarly, compared to TransUNet, our approach demonstrates gains of 4.14\%, 2.83\%, 2.97\%, and 3.22\%, respectively. When compared with recently established works like HiFormer \cite{hiformer} and Seg-Swinunet \cite{seg-swinunet}, we achieved 0.49\% and 0.18\% DSC improvement, respectively. This huge gain in DSC resulted from segmenting the difficult organs such as the Kidneys (L\&R) and Pancreas more accurately. The same is illustrated from Fig~\ref{synapse results}. We throw the light on Pancreas and Stomach segmentation highlighted in the \textcolor{yellow}{Yellow} (First row in Fig. \ref{synapse results}). Notably, SwinUNet could not segment either of them and other models \cite{transunet, levit, seg-swinunet} have not completely segmented the pancreas. The proposed model tackled this challenge very well and was on par with HiFormer.

\begin{table}[!ht]
    \centering
    \caption{Comparison of different methods in ACDC dataset. $\Delta_{UNet}$ denotes the improvement gain (\%) by comparing with the U-Net. $\Delta_{TransUNet}$ denotes the improvement gain (\%) by comparing with the TransUNet.}
    \label{tab:acdc}
    \resizebox{\columnwidth}{!}{
    \small
    \begin{tabular}{l|c|c|c|c} 
    \hline
    \textbf{Methods} & \textbf{Avg DSC} & \textbf{RV} & \textbf{Myo} & \textbf{LV} \\
    \hline
    UNet & 89.68 & 87.17 & 87.21 & 94.68 \\
    TransUNet & 89.71 & 86.67 & 87.27 & 95.18 \\
    SwinUNet & 88.07 & 85.77 & 84.42 & 94.03 \\
    LeViT-UNet & 88.21 & 85.56 & 84.75 & 94.32 \\
    Hiformer & 90.82 & 88.55 & 88.44 & 95.47 \\
    Seg-SwinUNet & 91.49 & 89.49 & 89.27 & 95.70 \\
    \hline
    \rowcolor{blue!20}
    STA-Unet & \textbf{92.25} & 90.31 & 90.44 & 95.99 \\
    \rowcolor{blue!20}
    $\Delta_{UNet}$ & \textcolor{blue}{+2.86} & \textcolor{blue}{+3.60} & \textcolor{blue}{+2.36} & \textcolor{blue}{+1.38}\\
    \rowcolor{blue!20}
    $\Delta_{TransUNet}$ & \textcolor{blue}{+2.83} & \textcolor{blue}{+4.19} & \textcolor{blue}{+3.63} & \textcolor{blue}{+0.85}\\    
    \hline
    \end{tabular}
    }
\end{table}

We established the Generalizability of our work by improving the DSC by 2.86\%, 6.53\%, and 6.03\% on ACDC, Glas, and MoNuSeg datasets, respectively, compared with UNet. We also outperformed all the Transformer based UNet architectures for segmentation tasks on ACDC and MoNuSeg and stood second best with a very minute difference of 0.64\% in DSC for Glas dataset (refer to Table. \ref{tab:acdc} \& \ref{tab:glas}). We visually compared the performance of Gland (using Glas) and Nuclear (using MoNuSeg) Segmentation in Fig. \ref{Glas results}. Poor foreground classification for Gland segmentation is clearly visible from the predictions of SwinUNet and LeViT-UNet (the top two rows in Fig~\ref{Glas results}), which has been remarkably tackled by the proposed Super Token Attention (STA-UNet) yielding accurate segmentation. We also highlight the difficulty distinguishing the foreground and background in the Glas dataset. In the case of MoNuSeg, our proposed model achieves results that are highly comparable to the ground truth, capturing complete shapes and maintaining clear backgrounds, even in challenging samples (as shown in the third row of Fig.~\ref{Glas results}). These findings reinforce our assertion that STA-UNet enhances segmentation performance, even with the reduced redundancy in shallow layer features typically seen in Transformer-based architectures.

\begin{table}[!t]
    \centering
    \caption{Comparison of different methods in Glas and MoNuSeg datasets. The second best performance is \underline{underlined}.}
    \label{tab:glas}
    \resizebox{\columnwidth}{!}{
    \Large
    \begin{tabular}{l|cc|cc}
    \hline
    \multicolumn{1}{c}{\multirow{2}{*}{\textbf{Method}}} & \multicolumn{2}{c}{\textbf{Glas}} & \multicolumn{2}{c}{\textbf{MoNuSeg}} \\
    \multicolumn{1}{c}{} & \multicolumn{1}{c}{ \textbf{DSC (\%)}} & \multicolumn{1}{c}{ \textbf{IOU (\%)}} & \multicolumn{1}{c}{ \textbf{DSC (\%)}} & \multicolumn{1}{c}{\textbf{IOU (\%)}}  \\
    \hline
    Unet  & 85.45$\pm$1.25 & 74.78$\pm$1.67 & 76.45$\pm$2.62 & 62.86$\pm$3.00 \\
    TransUNet & 88.40$\pm$0.74 & 80.40$\pm$1.04 & 78.53$\pm$1.06 & 65.05$\pm$1.28 \\
    SwinUNet & 89.58$\pm$0.57 & 82.06$\pm$0.73 & 77.69$\pm$0.94 & 63.77$\pm$1.15 \\
    LeViT-UNet & 81.19$\pm$1.38 & 69.73$\pm$1.85 & 70.28$\pm$3.92 & 53.08$\pm$0.43 \\
    Hiformer & 90.97$\pm$0.23 & 83.99$\pm$0.44 & 72.51$\pm$0.87 & 57.03$\pm$0.98 \\
    Seg-SwinUNet & \textbf{91.62$\pm$0.16}  & \textbf{85.29$\pm$0.30}  & \underline{79.38$\pm$0.15} & \underline{65.87$\pm$0.21} \\
    \hline
    STA-Unet & \underline{91.03$\pm$0.58} & \underline{84.29$\pm$0.94} & \textbf{81.06$\pm$0.66} & \textbf{68.24$\pm$0.80} \\
    \rowcolor{blue!20}
    $\Delta_{UNet}$ & \textcolor{blue}{+6.53} & \textcolor{blue}{+12.71} & \textcolor{blue}{+6.03} & \textcolor{blue}{+8.55}\\
    \rowcolor{blue!20}
    $\Delta_{TransUNet}$ & \textcolor{blue}{+2.97} & \textcolor{blue}{+4.84} & \textcolor{blue}{+3.22} & \textcolor{blue}{+4.90}\\ 
    \hline
    \end{tabular}
    }
\end{table}

\section{Conclusion}

In this study, we re-introduce Super Token Attention (STA) in UNet architecture as an STA Module to tackle the feature redundancy inherently present in existing transformer-based architectures while enhancing performance over organ segmentation tasks. We reported the preliminary analysis to mathematically make the redundancy present in Transformer-based architectures evident and encourage the research to mitigate the same. Our findings demonstrated a notable improvement over existing benchmarks across four publicly available datasets, evidencing the potential of STA-UNet in Medical Image Segmentation. Our extensive ablation study explains the impact on performance caused by two major parameter changes, i.e., Token size and Number of Attention heads. While our experiments are limited to the multi-organ segmentation tasks, the STA-UNet has the potential for broader applications such as Anomaly detection and restoration in various medical datasets. We anticipate further exploring the utility of the proposed architecture in other medical image applications in future research.

% In this study, we introduce the Super Token Attention (STA) module into the UNet architecture to address feature redundancy in Transformer-based models and enhance performance in organ segmentation tasks. Our preliminary analysis highlights the redundancy in existing Transformers and demonstrates how super tokens can mitigate this issue. We observed significant improvements over existing benchmarks on four publicly available datasets, underscoring STA-UNet's potential in medical image segmentation. Our ablation study assesses the effects of two key parameters—token size and number of attention heads—on performance. Although our experiments focus on medical image segmentation, STA-UNet shows promise for broader applications, such as anomaly detection and restoration in various medical datasets. Future research will explore these potential applications further.

% We introduce the Super Token Attention (STA) module in the UNet architecture to reduce feature redundancy in Transformer models and improve organ segmentation. Our results show significant performance gains on four datasets, highlighting STA-UNet’s effectiveness. We also explore the impact of token size and attention heads on performance. While our focus is on medical image segmentation, STA-UNet has potential for other medical applications, such as anomaly detection and restoration. Future research will explore these potential applications further.

%%%%%%%%% REFERENCES
{\small
\clearpage
\bibliographystyle{ieee_fullname}
\bibliography{egbib}
}

\end{document}